\documentclass{jpsj-suppl}
\usepackage{times}
\usepackage{color}
\title{
Spin-Orbit Interaction Effects in the Electronic Structure of B20-type CoSi: First-Principles Density Functional Study}
\author{Fumiyuki \textsc{Ishii}$^{1}$, Hiroki \textsc{Kotaka}$^{2}$, and Takashi \textsc{Onishi}$^{2}$}
\inst{$^{1}$Faculty of Mathematics and Physics, Kanazawa University, Kanazawa, 920-1192, Japan\\
$^{2}$Graduate School of Natural Science and Technology, Kanazawa University, Kanazawa, 920-1192, Japan
}
\email{fishii@mail.kanazawa-u.ac.jp}

\recdate{October 20, 2013}

\abst{
We have performed fully relativistic first-principles density functional calculations 
for non-magnetic B20-type CoSi.
The spin-orbit interaction 
has crucial effects on the electronic structures of a chiral crystal. 
The calculated band structure around the Fermi energy shows
Bloch vector $k$-linear dispersion expressed by 
a {\it real-spin} Weyl Hamiltonian, i.e., a mass-less Dirac Hamiltonian. 
We found the hedgehog-like spin textures in Bloch $\boldsymbol k$-vector space (momentum space) on the 
isoenergy surface around the $\Gamma$ point. 
The Fermi velocity for $k$-linear dispersion is about 0.22$v^g_F$,
where $v^g_F$ is the Fermi velocity of graphene.}

\kword{Spin-orbit interaction, Weyl point, Dirac cone, Spin texture, First-principles calculation}

\begin{document}
\maketitle

\section{Introduction}
B20-type monosilicides, such as $M$Si ($M$:transition metal), are known to show anomalous 
electrical and magnetic properties at low temperatures; examples include
a Kondo insulator in FeSi\cite{Degiorgi1999}  and 
a helical spin structure in MnSi\cite{shirane1983} and Fe$_{1-x}$Co$_x$Si\cite{Ishimoto1986}. 
The helical magnetic structures are thought to be due to a lack of an inversion center 
in its crystal structure. 
Then, the Dzyaloshinskii-Moriya interactions\cite{Dzyaloshinsky1958, Moriya1960} are induced by a relativistic 
spin-orbit interaction (SOI) effect. 
Recently, the spin-vortex, skyrmion-lattice phase was
discovered in the B20-type monosilicides\cite{Muehlbauer2009,Yu2010}
and the 
peculiar spin textures are expected to be applicable for spintronics applications\cite{Jonietz2010}.

In addition to the relativistic SOI effect on the spin textures in {\it real space}, 
there also exist relativistic SOI effects on the spin textures in the {\it momentum space}.
In nonmagnetic systems, with time-reversal symmetry, 
the breaking of the inversion symmetry of the crystal 
has a crucial effect on the electronic structures.
In general, the eigenenergies depends on the Bloch-states vector 
${\boldsymbol k}$ and spin vector ${\boldsymbol \sigma}$.
There is a relationship between the energies due to the time-reversal symmetry 
$E({\boldsymbol k},{\boldsymbol \sigma})=E(-{\boldsymbol k},-{\boldsymbol \sigma})$, as well as the 
space inversion symmetry $E({\boldsymbol k},{\boldsymbol \sigma})
=E(-{\boldsymbol k},{\boldsymbol \sigma})$.
Therefore, if the system has both symmetries, the electronic states at every 
${\boldsymbol k}$-point 
has doubly degenerate eigenenergies, as $E({\boldsymbol k},{\boldsymbol \sigma})
=E({\boldsymbol k},-{\boldsymbol \sigma})$.
On the other hand, if the system does not have space inversion symmetry, 
the eigenenergies have spin splitting, i.e., 
$E({\boldsymbol k},{\boldsymbol \sigma})\neq E({\boldsymbol k},-{\boldsymbol \sigma})$, except for 
the time-reversal invariant ${\boldsymbol k}$-points,
${\boldsymbol k=0, \boldsymbol G_{i}/2}$, where ${\boldsymbol G_{i}(i=1,2,3)}$ is the
reciprocal lattice vector.

This spin-splitting  is known for the structural inversion asymmetry (SIA) induced SOI effect. 
The Rashba effect\cite{Rashba1960} and Dresselhaus effect\cite{Dresselhaus1955} 
are well known for polar semiconductors and their heterostructures.
The ${\boldsymbol k}$-dependent spin Hamiltonian in noncentrosymmetric superconductors\cite{Frigeri2005, Frigeri2004a, Frigeri2004b} and  spin textures on the surface of materials\cite{Oguchi2009,Vajna2012} have been extensively studied in this context. Recently, the SIA-induced SOI effect has attracted much attention because it induces the current-driven spin transfer torque\cite{Miron2010}.
Although there have been several studies on the electronic structures of CoSi\cite{Imai2001,Sakai2007, Kudryavtsev2007}, the SOI effect has not been considered.

In this paper, we have performed 
fully relativistic first-principles density functional 
calculations 
for B20-type CoSi 
and discussed the effect of SOI in 
the electronic structure.
Using 
noncollinear spin-density functional theory implemented in 
{\scriptsize OPENMX} code\cite{OpenMX}, we investigated the spin texture and  electronic states.
We found that the spin-splitting $k$-linear band structures, 
Dirac cone bands,  
can be expressed by a 
{\it real-spin} Weyl Hamiltonian, i.e., $H_{eff}=v_F(k_x \sigma_x + k_y \sigma_y +k_z \sigma_z)$, where 
$v_F$ denotes the Fermi velocity.
The Dirac cone bands in CoSi originate from the B20 structure, 
chiral space group P2$_1$3.
As reported in a previous study, Dirac cone bands already exist in 
the scalar-relativistic (without SOI) band structure of 
CoSi\cite{Sakai2007}.
Such spin-degenerated Dirac cone bands are 
expressed by a {\it pseudo-spin}
Weyl Hamiltonian, i.e.,
{\it pseudo-spin} 
mass-less Dirac Hamiltonian
often discussed 
in the graphene. 
Recently, the group theoretical derivation of a {\it pseudo-spin} 
Weyl Hamiltonian for some space group has been reported\cite{Manes2012}.
Contrary to the {\it pseudo-spin} Weyl Hamiltonian, 
those of the {\it real-spin} discussed in this paper 
are quite important for
spin-dependent phenomena, as stated before.

\section{Methods}
Using the {\scriptsize OPENMX} code\cite{OpenMX},
we performed fully relativistic first-principles electronic-structure calculations based on density functional theory (DFT) within
the generalized gradient approximation (GGA)\cite{GGA-PBE}.
The norm-conserving pseudopotential method\cite{PP-TM} was used.
We used a linear combination of multiple pseudo atomic orbitals generated by a confinement scheme\cite{PAO1, PAO2}.
We used an (8,8,8) uniform k-point mesh. 
The pseudo atomic orbitals were expanded as follows: 
Co5.5-s3p3d3f1 and  Si5.5-s3p3d1.
The spin textures were calculated by post-processing calculation after 
the self-consistent field potential was obtained, as in our previous 
study\cite{Kotaka2013}. 
We calculated the $\boldsymbol k$-space spin density matrix 
$P_{\sigma, \sigma'}({\boldsymbol k},\mu)$ using the 
spinor wave-function whose component is 
given by $\Psi_{\sigma}({\boldsymbol r, \boldsymbol k},\mu)$ 
, obtained from the self-consistent calculations as follows: 
$P_{\sigma, \sigma'}({\boldsymbol k},\mu)=\int d{\boldsymbol r}
\Psi^{*}_{\sigma}({\boldsymbol r, \boldsymbol k},\mu) 
\Psi_{\sigma'}({\boldsymbol r, \boldsymbol k},\mu)$, 
where $\mu$ is the band index 
and $\sigma$ and $\sigma'$ 
are the spin indexes ($\uparrow, \downarrow$), respectively.
We deduced the spin polarization
in the $\boldsymbol k$-space from the 2 $\times$ 2 spin density matrix.
Because the wave function $\Psi_{\sigma}({\boldsymbol r, \boldsymbol k},\mu)$ 
is given by a linear combination of pseudo atomic orbitals, 
we can decompose the calculated spin polarization to its atomic components.
The crystal structure we used in this study is
the experimental lattice constant and internal parameters: 
$a$=4.444$\rm \AA$, $x_{\rm Co}$=0.143, 
$x_{\rm Si}$=0.844, for CoSi\cite{Teyssier2008}. 
In addition to the ($x$,$x$,$x$), 
there are three equivalent sites, 
3 permutations of ($x$+1/2, 1/2-$x$, $-x$).
Therefore, there are 8 atoms in the unitcell.

\section{Results and Discussions}
Figure 1 shows the calculated electronic band structure of CoSi.
Figure 1(a) shows a wide energy range covering 2 eV above/below the 
Fermi energy ($E_{\rm F}$). 
The calculated band structure of CoSi is similar to that
of scalar-relativistic (without SOI)
one\cite{Sakai2007}, as well as  
those of B20-type
FeSi\cite{Mattheiss1993}, 
MnSi\cite{Nakanishi1980,Jeong2004} and CoGe\cite{Kanazawa2012}.
Around the $E_{\rm F}$ and $\Gamma$ point, there are 
two types of characteristic dispersion bands.
 One is the Dirac cone-like $k$-linear dispersing bands (Dirac cone bands). 
The other one is  
much less dispersing bands (flat bands).
These characteristic band structures around the E$_F$, 
composed of the Dirac cone bands and the flat bands,  
are considered to be the origin for the large Seebeck 
coefficient\cite{Sakai2007, Kanazawa2012}.

Next, we focused on the spin splitting around the E$_F$.
In particular, we are interested in the Dirac cone bands; 
we plotted the band structure around the $E_{\rm F}$ 
on the $-{\rm M}(-0.5,-0.5,0)-{\rm \Gamma}(0,0,0)-{\rm M}(0.5,0.5,0)$ symmetry line, as shown in Fig. 1 (b).
The spin splitting of the bands can be seen,
except for the $\Gamma$ point 
due to 
the SIA-induced SOI effect mentioned above.
Here, we speculate the symmetry character of the bands 
at the $\Gamma$-point around the $E_{\rm F}$
based on the known 
double group character table\cite{Bradley1972}.
Without considering SOI, 
the $E_{\rm F}$ lies at the spin degenerate orbital triplet state $\Gamma_4$. 
Because of the SOI, the 6-fold degenerate $\Gamma_4$ is split into 
a lower-energy doublet and two higher-energy doublets. 
In other words, around below the $E_{\rm F}$ at the $\Gamma$-point, 
there is a $\Gamma_5$ doublet.
$\Gamma_6$ and $\Gamma_7$, two degenerate doublets, are above the $E_{\rm F}$. 
The spin-orbit splitting of CoSi at the $\Gamma$ point is about 54.6 meV.
The Dirac cone bands cross above the $E_{\rm F}$. 
Then the so-called 
Weyl point is in the degenerate doublets, $\Gamma_6$ and $\Gamma_7$.

We evaluated the Fermi velocity of the Dirac cone bands. 
The obtained Fermi velocity for one of the Dirac cone bands, 
the higher 
energy band, is about 0.195 $\times10^6$ m/s, which is about
22$\%$ that of graphene, 0.885 $\times10^6$ m/s.
To proceed with the analysis of the Dirac cone bands, we extracted the 
Dirac cone band as shown in Fig. 2(a). We also plotted the calculated 
spin polarization $\boldsymbol P$(P$_x$, P$_y$, P$_z$) as a vector arrow 
on the bands. Because the $-{\rm M}-{\rm \Gamma}-{\rm M}$ line lies in 
the $k_xk_y$ plane, we projected the $P_x$ and $P_y$ component on the line 
(horizontal axis) and plotted the $P_z$ component on the vertical axis.
The horizontal arrows in Fig. 2 (a) mean that 
there is no $P_z$ component. 
This result is consistent with the Dirac cone bands expressed by 
the {\it real-spin} Weyl Hamiltonian, i.e., $H_{eff}=v_F(k_x \sigma_x + k_y \sigma_y +k_z \sigma_z)$. 

Figure 2 (b) shows the band structure around the $\Gamma$-point above 
the Weyl point. There are two bands; one band is 
the linear dispersing Dirac cone band, and 
the other is a relatively parabolic band.
The parabolic band is connected to the flat band
below the band crossing point.
Here, we are interested in the differences in the magnitude of spin polarization between 
the Dirac cone band and parabolic band. As seen in Fig. 2(b), the magnitude of the spin polarization of the parabolic band decreases when 
it is close to the degenerate point. On the other hand, as seen in Fig. 2(a) and (b), the magnitude of the spin polarization in the Dirac cone band is almost constant. The magnitude of the spin polarization of the Dirac cone band is twice as large as that of the parabolic band for the states at 0.06 eV above the $E_{\rm F}$, as shown in Fig. 2 (c).
We decomposed the spin polarization to the atomic components. 
As a result, at states 0.06 eV above the $E_{\rm F}$, 
the spin polarization consists of more than 
80 percent Co orbitals
for both 
Dirac cone and parabolic bands.

We show the hedgehog-like spin texture on the isoenergy for 0.2 eV above the Fermi energy in Fig. 3.
The spin texture of the lower-energy parabolic dispersion 
shows spin directed from the $\Gamma$ point outwards, 
as shown in Fig. 3(a). 
On the other hands, the spin texture of the higher-energy 
linear dispersion shows spin 
directed towards the 
$\Gamma$ point, as shown in Fig. 3(b).
These results indicate that 
both the Dirac cone bands and 
parabolic bands are expressed by the   
{\it real-spin} Weyl Hamiltonian.

The origin of hedgehog-like spin texture 
is the chiral crystal structure. 
Space group P2$_1$3 does not have mirror symmetry, 
so the spin Hamiltonian, which is linear in $k$, will have 
a nonzero $k_i\sigma_i(i=x,y,z)$ term. 
If we have $m_{xy}$ mirror symmetry where the mirror plane lies in the $xy$ plane, 
the polar vector $\boldsymbol{k}(k_x, k_y, k_z$) is transformed to ($k_x, k_y, -k_z$) and 
the axial vector $\boldsymbol{\sigma}(\sigma_x, \sigma_y, \sigma_z$) is transformed to ($-\sigma_x, -\sigma_y, \sigma_z$). 
Then, the $k_i\sigma_i(i=x,y,z)$ terms disappear in 
the non-chiral crystal structure.
There are  twelve symmetry operations in space group P2$_1$3, 
\{E, 4C$_3$, 4C$_3^2$, 3C$_2$ \}. 
We can derive the Weyl Hamiltonian from the rotational invariant 
condition for the Hamiltonian. Thus, in space group P2$_1$3, $k_i\sigma_i(i=x,y,z)$ terms only remain in the Hamiltonian.
The spin Hamiltonian can also be derived from group theory
\cite{Oguchi2009,Vajna2012, Frigeri2005,Samokhin2009}.
Our derivation of the spin Hamiltonian is simpler than those given in 
previous studies.
\section{Summary}
We have performed fully relativistic first-principles density functional calculations for non-magnetic B20-type CoSi with SOI. 
The calculated band structures around the Fermi energy show $k$-linear dispersion expressed by 
a real-spin Weyl Hamiltonian, i.e., mass-less Dirac Hamiltonian. 
We found the hedgehog-like spin textures in momentum space on the 
isoenergy surface around the $\Gamma$ point. 
The Fermi velocity for the linear dispersion is about 0.22$v^g_F$,
where $v^g_F$ is the Fermi velocity of graphene.
Our findings
provide the basis for the 
further studies
on noncentrosymmetric superconductivity\cite
{Frigeri2004a, Frigeri2004b} 
and spin related transport properties\cite{Ohta1998,Onose2005} in B20-type materials.

\section*{Acknowledgements}
The authors thank the Yukawa Institute for Theoretical Physics at Kyoto University. 
Discussions during the YITP workshop YITP-W-13-01 on 
"Dirac electrons in solids" were useful to complete this work.
Part of this research has been funded by the MEXT HPCI Strategic Program. 
This work was partly supported by Grants-in-Aid for Scientific Research 
(Nos. 25104714, 25790007, and 25390008) from the JSPS.
The computations in this research were performed using 
the supercomputers at the ISSP, the University of Tokyo.
\begin{figure}
\begin{center}
\includegraphics[width=16cm]{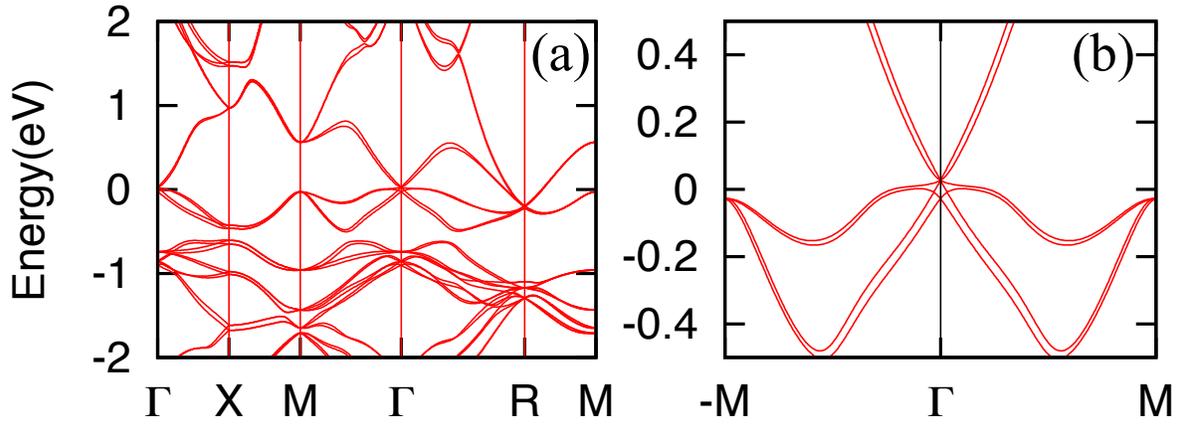}
\end{center}
\caption{(Color online) 
Fully relativistic calculated band structure of CoSi, 
a wide energy range covering 2 eV above/below the Fermi energy (a)
and around the Fermi energy (b).
The origin of the energy is taken at the Fermi energy.
The high-symmetry ${\boldsymbol k}$-point symbols in the first Brillouin zone are denoted as 
$\rm \Gamma$(0,0,0), X(0.5,0,0), M(0.5,0.5,0), R(0.5,0.5,0.5), -M(-0.5,-0,5,0)
by the $\frac{\pi}{a}$ unit. 
}
\label{f1}
\end{figure}

\begin{figure}
\begin{center}
\includegraphics[width=16cm]{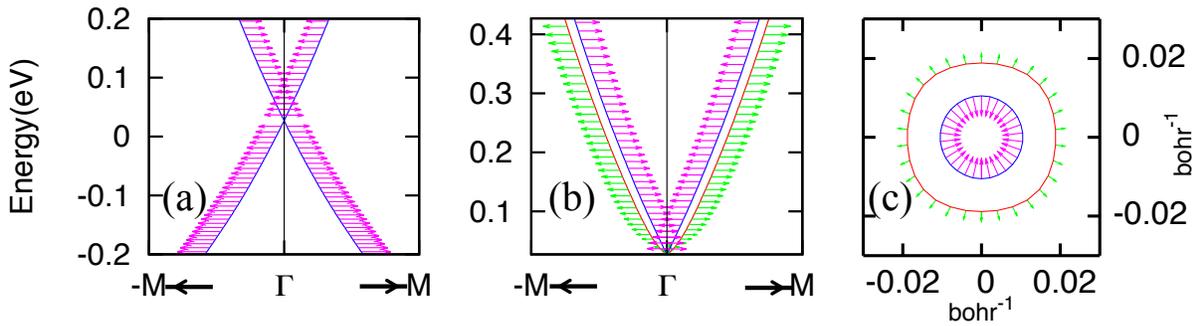}
\end{center}
\caption{(Color online) 
Fully relativistic calculated band structure around the $\Gamma$ point for CoSi with calculated 
spin polarization. (a) The band structure of the extracted $k$-linear dispersion, (b) 
the band structure above the Weyl point. (c) The isoenergy (0.06eV) 
line with calculated spin polarization on 
$k_x-k_y$ plane ($k_z=0$).
}
\label{f2}
\end{figure}

\begin{figure}
\begin{center}
\includegraphics[width=16cm]{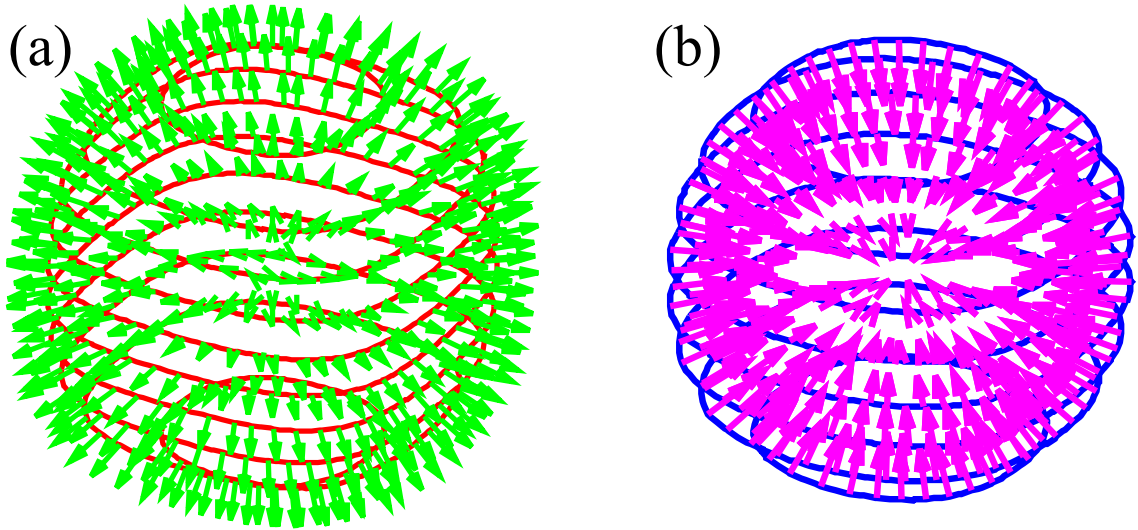}
\end{center}
\caption{(Color online) 
Calculated spin polarization on the isoenergy surface 
for 0.2eV above the Fermi energy for 
(a) the parabolic band and (b) the Dirac cone band.
}
\label{f3}
\end{figure}


\begin{thebibliography}{10}

\bibitem{Degiorgi1999}
L.~Degiorgi: Rev. Mod. Phys. {\bfseries 71} (1999) 687.

\bibitem{shirane1983}
G.~Shirane, R.~Cowley, C.~Majkrzak, J.~Sokoloff, B.~Pagonis, C.~Perry, and
  Y.~Ishikawa: Phys. Rev. B {\bfseries 28} (1983) 6251.

\bibitem{Ishimoto1986}
K.~Ishimoto, Y.~Yamaguchi, S.~Mitsuda, M.~Ishida, and Y.~Endoh: J. Magn. Magn.
  Mater. {\bfseries 54} (1986) 1003.

\bibitem{Dzyaloshinsky1958}
I.~Dzyaloshinsky: J. Phys. Chem. Solids {\bfseries 4} (1958) 241.

\bibitem{Moriya1960}
T.~Moriya: Phys. Rev. {\bfseries 120} (1960) 91.

\bibitem{Muehlbauer2009}
S.~Muehlbauer, B.~Binz, F.~Jonietz, C.~Pfleiderer, A.~Rosch, A.~Neubauer,
  R.~Georgii, and P.~Boeni: Science {\bfseries 323} (2009) 915.

\bibitem{Yu2010}
X.~Z. Yu, Y.~Onose, N.~Kanazawa, J.-H. Park, J.~H. Han, Y.~Matsui, N.~Nagaosa,
  and Y.~Tokura: Nature {\bfseries 465} (2010) 901.

\bibitem{Jonietz2010}
F.~Jonietz, S.~M{\"u}hlbauer, C.~Pfleiderer, A.~Neubauer, W.~M{\"u}nzer,
  A.~Bauer, T.~Adams, R.~Georgii, P.~B{\"o}ni, R.~A. Duine, K.~Everschor,
  M.~Garst, and A.~Rosch: Science {\bfseries 330} (2010) 1648.

\bibitem{Rashba1960}
E.~I. Rashba: Sov. Phys. Solid State {\bfseries 2} (1960) 1109.

\bibitem{Dresselhaus1955}
G.~Dresselhaus: Phys. Rev. {\bfseries 100} (1955) 580.

\bibitem{Frigeri2005}
P.~A. Frigeri: Doctoral and Habilitation Theses, ETH Z{\"u}rich  (2005).

\bibitem{Frigeri2004a}
P.~A. Frigeri, D.~F.~D. Agterberg, A.~A. Koga, and M.~M. Sigrist: Phys. Rev.
  Lett. {\bfseries 92} (2004) 097001.

\bibitem{Frigeri2004b}
P.~A. Frigeri, D.~F. Agterberg, A.~Koga, and M.~Sigrist: Phys. Rev. Lett.
  {\bfseries 93} (2004) 99903.

\bibitem{Oguchi2009}
T.~Oguchi and T.~Shishidou: J. Phys.: Condens. Matter {\bfseries 21} (2009)
  2001.

\bibitem{Vajna2012}
S.~Vajna, E.~Simon, A.~Szilva, K.~Palotas, B.~Ujfalussy, and L.~Szunyogh: Phys.
  Rev. B {\bfseries 85} (2012) 75404.

\bibitem{Miron2010}
I.~M.~Miron, G.~Gaudin, S.~Auffret, B.~Rodmacq, A.~Schuhl, S.~Pizzini,
  J.~Vogel, and P.~Gambardella: Nature Materials {\bfseries 9} (2010) 230.

\bibitem{Imai2001}
Y.~Imai, M.~Mukaida, K.~Kobayashi, and T.~Tsunoda: Intermetallics {\bfseries 9}
  (2001) 261.

\bibitem{Sakai2007}
A.~Sakai, F.~Ishii, Y.~Onose, Y.~Tomioka, S.~Yotsuhashi, H.~Adachi, N.~Nagaosa,
  and Y.~Tokura: J. Phys. Soc. Jpn. {\bfseries 76} (2007) 093601.

\bibitem{Kudryavtsev2007}
Y.~V. Kudryavtsev, V.~A. Oksenenko, Y.~P. Lee, J.~Y. Rhee, and Y.~D. Kim: J.
  Appl. Phys. {\bfseries 102} (2007) 3503.

\bibitem{OpenMX}
T.~Ozaki, H.~Kino, J.~Yu, M.~J. Han, M.~Ohfuti, F.~Ishii, K.~Sawada, Y.~Kubota,
  T.~Ohwaki, H.~Weng, M.~Toyoda, H.~Kawai, Y.~Okuno, R.~Perez, P.~P. Bell,
  T.~Duy, Y.~Xiao, A.~M. Ito, and K.~Terakura: Available from:
  (http://www.openmx-square.org/) .

\bibitem{Manes2012}
J.~L. Ma{\~n}es: Phys. Rev. B {\bfseries 85} (2012) 155118.

\bibitem{GGA-PBE}
J.~P. Perdew, K.~Burke, and M.~Ernzerhof: Phys. Rev. Lett. {\bfseries 77}
  (1996) 3865.

\bibitem{PP-TM}
N.~Troullier and J.~L. Martins: Phys. Rev. B {\bfseries 43} (1991) 1993.

\bibitem{PAO1}
T.~Ozaki: Phys. Rev. B {\bfseries 67} (2003) 155108.

\bibitem{PAO2}
T.~Ozaki and H.~Kino: Phys. Rev. B {\bfseries 69} (2004) 195113.

\bibitem{Kotaka2013}
H.~Kotaka, F.~Ishii, and M.~Saito: Jpn. J. App. Phys. {\bfseries 52} (2013)
  5204.

\bibitem{Teyssier2008}
J.~Teyssier, R.~Viennois, J.~Salamin, E.~Giannini, and D.~Van Der~Marel: J.
  Alloys Compd. {\bfseries 465} (2008) 462.

\bibitem{Mattheiss1993}
L.~F. Mattheiss and D.~R. Hamann: Phys. Rev. B {\bfseries 47} (1993) 78100.

\bibitem{Nakanishi1980}
O.~Nakanishi, A.~Yanase, and A.~Hasegawa: J. Magn. Magn. Mater. {\bfseries 15}
  (1980) 879.

\bibitem{Jeong2004}
T.~Jeong and W.~E. Pickett: Phys. Rev. B {\bfseries 70} (2004) 075114.

\bibitem{Kanazawa2012}
N.~Kanazawa, Y.~Onose, Y.~Shiomi, S.~Ishiwata, and Y.~Tokura: Appl. Phys. Lett.
  {\bfseries 100} (2012) 3902.

\bibitem{Bradley1972}
C.~J. Bradley and A.~P. Cracknell: {\it The mathematical theory of symmetry in
  solids}  (Clarendon Press, Oxford, 2007).

\bibitem{Samokhin2009}
K.~V. Samokhin: Ann. Phys. {\bfseries 324} (2009) 2385.

\bibitem{Ohta1998}
H.~Ohta, T.~Arioka, E.~Kulatov, S.~Mitsudo, and M.~Motokawa: J. Magn. Magn.
  Mater. {\bfseries 177} (1998) 1371.

\bibitem{Onose2005}
Y.~Onose, N.~Takeshita, C.~Terakura, H.~Takagi, and Y.~Tokura: Phys. Rev. B
  {\bfseries 72} (2005) 224431.

\end{thebibliography}

\end{document}